\documentclass[aps,prc,tightenlines,floatfix,showpacs,twocolumn]{revtex4}
\usepackage{graphics}
\usepackage{bm}
\usepackage{amsmath}
\usepackage{amssymb}
\usepackage{txfonts} 
\usepackage{supertabular}
\usepackage{array}                    

\newcommand{\Be}{Be}
\newcommand{\He}{He}

\begin{document}

\title{Scattering of nucleons from nuclei with couplings to 
particle-unstable excited states.}

\author{P. R. Fraser$^{1,2}$}
\email{paul.fraser@nucleares.unam.mx}
\author{K. Amos$^{2}$}
\author{L. Canton$^{3}$}
\author{S. Karataglidis$^{4}$}
\author{J. P. Svenne$^{5}$}
\author{D. van der Knijff$^{6}$}

\affiliation{$^{1}$ Instituto de Ciencias Nucleares, Universidad
  Nacional Aut\'onoma de M\'exico, 04510 M\'exico, D.F., Mexico}
\affiliation{$^{2}$ School of Physics, University of Melbourne,
  Victoria 3010, Australia} 
\affiliation{$^{3}$ Istituto Nazionale di Fisica Nucleare, Sezione di
  Padova, I-35131, Italy}
\affiliation{$^{4}$ Department of Physics,
  University of Johannesburg, South Africa}
\affiliation{$^{5}$
  Department of Physics and Astronomy, University of Manitoba and
  Winnipeg Institute for Theoretical Physics, Winnipeg, MB, R3T 2N2,
  Canada,}
\affiliation{$^{6}$ Advanced Research Computing, Information
  Division, University of Melbourne, Victoria 3010, Australia}

\date{\today}

\begin{abstract}
  The physics of radioactive ion beams implies the description of
  weakly-bound nuclear systems. One key aspect concerns the coupling
  to low-lying collective-type excited states, which for these systems
  might not be stable levels, but particle emitting resonances. In
  this work we describe how the scattering cross section and compound
  spectra change when the colliding fragments have such collective
  excitations featuring particle emission. We explore this question in
  the framework of a multi-channel algebraic scattering method of
  determining nucleon-nucleus cross sections at low energies. For a
  range of light-mass, particle-unstable nuclear targets, scattering
  cross sections as well as the spectra of the compound nuclei formed
  have been determined from calculations that do and do not consider
  particle emission widths for nuclear states. Assuming a resonance
  character for target states markedly varies evaluated cross sections
  from those obtained assuming the target spectrum to have entirely
  discrete states.
\end{abstract}


\maketitle

\section{Introduction}
\label{Sec-Introduction}

The availability of radioactive ion beams (RIB) at many laboratories
has given experimental information on many exotic nuclei. Those
studies have revealed novel structures, such as nuclear skins and
halos.  Of particular interest are data obtained from scattering
exotic nuclei from hydrogen targets, which equates in inverse
kinematics to proton scattering from those nuclei. For incident
energies of many tens of MeV per nucleon, such data have been analyzed
in terms of effective nucleon-nucleus interactions used in distorted
wave approximations \cite{Am00,La01,St02}. For lower energies, at
which the states in the low excitation regime of the target are
influential, coupled-channels approaches~\cite{Ri98,Am03} are needed.
Coupled-channel analyses of RIB data made so far assumed nuclei to
have spectra of discrete (zero-width) states. There has been a study
of such systems at intermediate energies which sought to account for
the low binding by including additional decay
channels~\cite{Go09}. However, those authors did not consider decay
widths for the nuclear states included in those coupled channels.
Radioactive nuclei, and especially those near the drip lines, have
quite low particle emission thresholds and thus can have low-lying
resonance states in their spectra.

This paper considers nucleon scattering from a range of weakly-bound,
light-mass nuclei, for which discrete resonance effects in the elastic
cross section are present. Evaluations of both the scattering cross sections 
and of the spectra of the compound nucleus have been made using the
development~\cite{Fr08a} in which particle emission widths are accounted for 
within a multi-channel algebraic scattering (MCAS) theory~\cite{Am03}. 
With MCAS, solutions of coupled
Lippmann-Schwinger equations are found (in momentum space) by using
finite-rank separable representations of an input matrix of
nucleon-nucleus interactions. The separable form factors are generated
from an set of sturmian functions~\cite{We65}. Those sturmian functions
are generated from the same input matrix of interactions.
Details are given in Refs.~\cite{Am03,Ca05}.

There are distinct advantages in using the MCAS method.  It includes a
prescription~\cite{Am03} that allows one to locate all compound-system
resonance centroids and widths regardless of how narrow they may
be. Further, by use of orthogonalizing pseudo-potentials (OPP) in
generating the sturmians and solving the Lippmann-Schwinger equations,
it is ensured that the Pauli principle is not
violated~\cite{Ca05}. That is so even if a collective model is used to
specify nucleon-nucleus interactions. The latter is of
paramount importance for coupled-channel calculations~\cite{Am05}, as
otherwise some compound nucleus wave functions so defined possess
spurious components.

MCAS has been used previously to characterize scattering from nuclei
away from the valley of stability~\cite{Ca06,Ca06a}, but in those
studies, the target nuclei states were all taken to be of zero
width. In this study, several of those previous investigations are
amended to include particle-unstable channel widths.

In Section~\ref{Sec-Theory}, a brief description is given of the MCAS
method and how resonant character of a target state effects it. Then
in Sections~\ref{Sec-Be8+n} through \ref{Sec-He6+n}, examples are
described and results presented.  A discussion is given, and
conclusions drawn, in Section~\ref{conclus}.

\section{Theoretical development}
\label{Sec-Theory}

The MCAS method was developed to find solutions of coupled-channel, 
partial-wave expanded, Lippmann-Schwinger equations for 
the scattering of two nuclei. For each total system spin-parity
($J^\pi$), those equations are
\begin{multline}
T^{J^{\pi}}_{cc'}( p, q; E ) = 
  V^{J^{\pi}}_{cc'}( p, q )  \\ 
+ \mu  \left[ \sum^{\text{open}}_{c'' = 1} \int^{\infty}_0
V^{J^{\pi}}_{cc''}( p, x) \frac{ x^2 }{ k^2_{c''} - x^2 + i\varepsilon
  } T^{J^{\pi}}_{c''c'}( x, q; E ) \, dx \right. 
 \\
 - \left. \sum^{\text{closed}}_{c'' = 1} \int^{\infty}_0
  V^{J^{\pi}}_{cc''}( p, x ) \frac{ x^2 }{ h^2_{c''} + x^2 }
  T^{J^{\pi}}_{c''c'}( x,q; E ) \, dx \right] \, ,
\label{LScc'}
\end{multline}
where the channels are denoted $c$ and where $\mu
=\textstyle\frac{2\overline{m}}{\hbar}$, ${\overline m}$ being the
reduced mass. There are two summations as the open and closed-channel
components are separated, with wave numbers
\begin{equation}
k_c = \sqrt{\mu(E - \epsilon_c)} \: \text{\ and\ } \: h_c = 
\sqrt{\mu(\epsilon_c - E)}\; ,
\label{kandh}
\end{equation}
for $E > \epsilon_c$ and $E < \epsilon_c$ respectively. $\epsilon_c$
is the energy threshold at which channel $c$ opens. In the nucleon
scattering case, they coincide with the, presumed discrete,  excitation 
energies of the nucleus.  Henceforth the $J^\pi$ superscript 
is to be understood. Expansion of $V_{cc'}$ in terms of a finite 
number ($N$) of sturmians leads to an algebraic representation of the 
scattering matrix~\cite{Am03}
\begin{multline}
S_{cc'} =\; \delta_{cc'} \; -\, i^{(l_{c'} - l_c + 1)}\; \pi\; \mu 
\\
\times 
\sum^{N}_{n,n'=1} 
\sqrt{k_c} \hat{\chi}_{cn}(k_c) \left( \left[ {\mbox{\boldmath
$\eta$}} - \mathbf{G}_0 \right]^{-1} \right)_{nn'} 
  \hat{\chi}_{c'n'} (k_{c'})
  \sqrt{k_{c'}}\; .
  \label{Smatrix}
\end{multline}
The indices $c$ and $c'$ refer now only to open channels, $l_c$ is the
partial wave in channel $c$ and the Green's function matrix is
\begin{multline}
  \left[ \mathbf{G}_0 \right]_{nn'} = \mu \left[
  \sum^{\text{open}}_{c = 1} \int^{\infty}_0 \hat{\chi}_{cn}(x)
  \frac{x^2}{ k^2_c - x^2 + i\varepsilon} \hat{\chi}_{cn'}(x) \, dx
  \right. \\
   - \left. \sum^{\text{closed}}_{c = 1}
  \int^{\infty}_{0} \hat{\chi}_{cn}(x) \frac{ x^2 }{ h^2_c + x^2 }
  \hat{\chi}_{cn'}(x) \, dx \right].
  \label{S-Gmod}
\end{multline}
Here, \mbox{\boldmath $\eta$}  is a column vector of sturmian eigenvalues 
and $\hat \chi$ are form factors determined from the chosen sturmian
functions. Details are given in Ref.~\cite{Am03}.

Traditionally, all target states are taken to have eigenvalues of zero
width, and the (complex) Green's functions are evaluated using the
method of principal parts. This assumes time evolution of target
states is given by
\begin{equation}
\left|x,t\right\rangle = e^{-i H_0 t/\hbar}\; \left|x,t_0\right\rangle = 
e^{-i E_0 t/\hbar}\; \left|x, t_0 \right\rangle.
\end{equation}
However, if states decay, they evolve as
\begin{equation}
\left|x,t\right\rangle = e^{-\textstyle\frac{\Gamma}{2}t} \; e^{-iE_0t/\hbar}\; 
\left|x,t_0\right\rangle.
\end{equation}

Thus, in the Green's function, channel energies are complex, 
as are the squared channel wave numbers,
\begin{equation}
\hat{k_c}^2 = \mu \left(E - \epsilon_c + \textstyle\frac{i\Gamma_c}{2} 
\right) \: \: {\rm and}  \:\:
\hat{h_c}^2 = \mu \left( \epsilon_c - E - \textstyle\frac{i\Gamma_c}{2}
\right) .
\label{hatkandhath}  
\end{equation}  
$\textstyle\frac{ \Gamma_c }{2}$ is half the width of the target  state 
associated with channel $c$. The Green's function matrix 
elements then are
\begin{multline}
  \left[ \mathbf{G}_0 \right]_{nn'}  = \mu \left[
  \sum^{\text{open}}_{c = 1} \int^{\infty}_0 \hat{\chi}_{cn}(x)
  \frac{x^2}{ k_c^2 - x^2 + \textstyle\frac{ i \mu \Gamma_c }{2} } 
  \hat{\chi}_{cn'}(x) \, dx
  \right. \\
  - \left. \sum^{\text{closed}}_{c = 1}
  \int^{\infty}_{0} \hat{\chi}_{cn}(x) 
  \frac{ x^2 }{ h_c^2 + x^2 - \textstyle\frac{ i \mu \Gamma_c }{2} }
  \hat{\chi}_{cn'}(x) \, dx \right] ,
  \label{S-Ggamma2}
\end{multline}
where $k_c$ and $h_c$ are as in Eq.~(\ref{kandh}).  With poles moved
significantly off the real axis, direct integration of a complex
integrand along the real momentum axis is feasible.  This has been
done. However, for any infinitesimal-width target state, or resonance
so narrow that it can be treated as such, the method of principal
parts has been retained.

An additional complication arises from the Lorentzian shape of target
resonances as modeled by MCAS. Below threshold ($E=0$), resonance
(target state) widths must not cause instability for any compound
system that has subthreshold states.  To ensure this, an energy
dependent scaling factor must be applied to target resonance
states. As a first, minimal correction, this investigation applies a
Heaviside step function at $E=0$, forcing the target state spread to
vanish below threshold.  Eq.~(\ref{S-Ggamma2}) then becomes
\begin{multline}
  \left[ \mathbf{G}_0 \right]_{nn'}  
  = H(E) \cdot \mu \left[
  \sum^{\text{open}}_{c = 1} \int^{\infty}_0 \hat{\chi}_{cn}(x)
  \frac{x^2}{ k_c^2 - x^2 + \textstyle\frac{ i \mu \Gamma_c }{2} } 
  \hat{\chi}_{cn'}(x) \, dx
  \right. \\
  \hspace{2.2cm}- \left. \sum^{\text{closed}}_{c = 1}
  \int^{\infty}_{0} \hat{\chi}_{cn}(x) 
  \frac{ x^2 }{ h_c^2 + x^2 - \textstyle\frac{ i \mu \Gamma_c }{2} }
  \hat{\chi}_{cn'}(x) \, dx \right]\\
  + (1-H(E)) \cdot \mu \left[
  \sum^{\text{open}}_{c = 1} \int^{\infty}_0 \hat{\chi}_{cn}(x)
  \frac{x^2}{ k^2_c - x^2 } \hat{\chi}_{cn'}(x) \, dx
  \right. \\
   - \left. \sum^{\text{closed}}_{c = 1}
  \int^{\infty}_{0} \hat{\chi}_{cn}(x) \frac{ x^2 }{ h^2_c + x^2 }
  \hat{\chi}_{cn'}(x) \, dx \right].
  \label{S-GgammaH}
\end{multline}

In practice, however, this is not sufficient, as
reaction cross sections approaching threshold increase asymptotically
(see Section~\ref{Sec-Be8+n}). More sophisticated methods to address
this issue are in development, and indeed, one aim of this paper is to
emphasize the need for such improvement.

\section{$n$+$^{8}$\protect\Be\:  scattering}
\label{Sec-Be8+n}

The low excitation $^8$Be spectrum has a $0^+;0$ ground state that has
a small width for decay into two $\alpha$-particles ($6 \times
10^{-6}$ MeV), a broad $2^+;0$ resonance state with centroid at 3.03
MeV and width of 1.5 MeV, followed by a broader $4^+;0$ resonance
state with centroid at 11.35 MeV and width ${\sim}3.5$
MeV~\cite{Ti04}, both of which also decay into $\alpha$-particles.  As
the width of the ground state is so small, it is approximated to be a
zero-width state in our MCAS calculations.

\begin{figure}[htp]
\begin{center}
\scalebox{0.53}{\includegraphics*{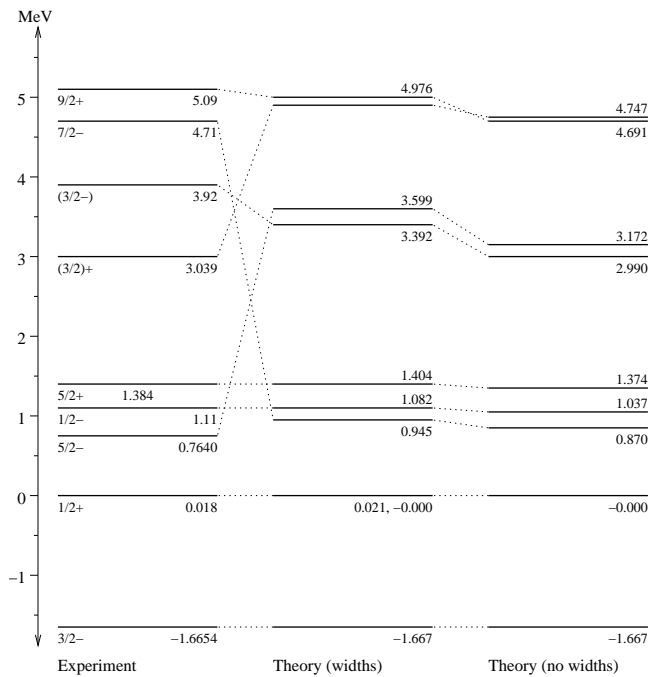}}
\end{center}
\caption{ \label{Be8+n-spec}
The experimental $^9$Be spectrum and those calculated from 
MCAS evaluations of the neutron+$^8$Be system
assuming the ${}^8$Be states are resonances (widths) 
or are discrete (no widths).} 
\end{figure}

Two evaluations of the $n$+$^8$Be cross section have been obtained
with MCAS. In the first, both the $2^+$ and $4^+$ states were also
taken to have zero-width and, as will be used throughout, results from
such calculations are identified by the label `no-width(s)'.  In the
second calculation, the true widths as tabulated~\cite{Ti04} were used
and results of such hereafter are identified by the label `width(s)'.
In both calculations, the same nuclear interaction was used. It was
specified from a collective model of the nucleus with rotor
character~\cite{Am03}.  The parameter values used were chosen to give
a fit to the low lying spectrum of ${}^9$Be, and they are listed in
Table~\ref{Be8-param}.  They differ from those used in the previous
calculations~\cite{Fr08a}, as some aspects of the experimentally
determined structure of $^9$Be are better reproduced.  The spectra
from the two calculations are depicted in Fig.~\ref{Be8+n-spec}.
While some disparities remain, to be expected from the simple
prescription we have used to specify the interaction potentials,
compared to the previous results~\cite{Fr08a} the centroid energy of
the ${\textstyle\frac{1}{2}}^+$ first resonance is now placed above threshold in
the widths case (see later regarding the degeneracy), and there are
now no spurious states in the 0 to 5.5 MeV projectile energy range.
The definition of all parameters can be found in Ref.~\cite{Am03},
with the addition of values for Pauli hindrance for the $0s_{1/2}$ and
$0p_{3/2}$ subshells of the target states, as defined in
Ref.~\cite{Ca06, Ca06a}.

\begin{table}\centering
\caption
{\label{Be8-param} 
Parameter values defining the nucleon-$^8$Be interaction.}
\begin{supertabular}{>{\centering}p{28mm} p{27mm}<{\centering} 
p{27mm}<{\centering}}
\hline
\hline
 & Odd parity & Even parity\\
\hline
$V_{\rm central}$ (MeV) & -31.5 & -42.2 \\
$V_{l l}$ (MeV)       & 2.0   & 0.0 \\
$V_{l s}$ (MeV)       & 12.0 & 11.0 \\
$V_{ss}$ (MeV)        & -2.0 & 0.0 \\
\end{supertabular}
\begin{supertabular}{>{\centering}p{19.5mm} p{21mm}<{\centering} 
>{\centering}p{21mm} p{19mm}<{\centering} }
\hline
\hline
Geometry & $R_0 = 2.7$ fm & $a = 0.65$ fm & $\beta_2 = 0.7$ \\ 
Coulomb$^\dag$ &  $R_c = 1.3$ fm & $a_c = 1.00$ fm  & $\beta_2 = 0.7$ \\
\end{supertabular}
\begin{supertabular}{>{\centering}p{28mm} p{27mm}<{\centering} 
p{27mm}<{\centering}}
\hline
\hline
                        &$0s_{1/2}$ &$0p_{3/2}$\\
\hline
 $0^+ \; \lambda^{(OPP)}$ & 1000 & 0.0 \\
 $2^+ \; \lambda^{(OPP)}$ & 2.0 & 0.2 \\
 $4^+ \; \lambda^{(OPP)}$ & 0.0 & 0.0 \\
\hline
\hline
\end{supertabular}
$^\dag$ for proton-${}^8$Be calculations (see Section~\ref{Sec-Be8+p}).
\end{table}

\begin{figure}[htp]
\begin{center}
\scalebox{0.37}{\includegraphics*{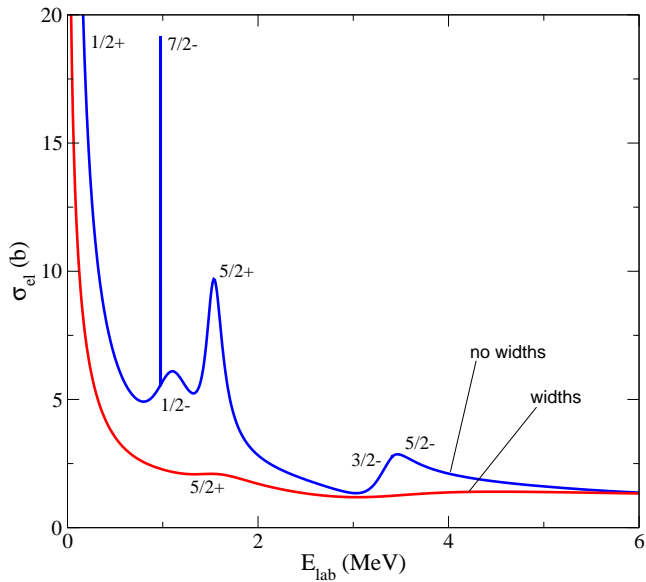}}
\end{center}
\caption{ \label{Be8+n}
(Color online) Calculated cross sections for neutron scattering from 
$^8$Be found using MCAS with and without the resonance character
in the target states.}
\end{figure}

Note that the ${\textstyle\frac{1}{2}}^+$ state is naturally moved from the
positive to negative regime with the removal of widths. Thus, the
application of the Heaviside function means that in the widths case
there are two representations of this state; the one frozen below zero
and that moved above zero. This phenomenon is further discussed in
Section~\ref{Sec-He6+p}.

The results for the scattering cross sections are shown in
Fig.~\ref{Be8+n}.  Upon introducing target-state widths, as found
previously ~\cite{Fr08a}, the resonances linked to spectral properties
of $^9$Be are suppressed. They are still present but their widths have
increased and their magnitudes decreased so as only the
${\textstyle\frac{5}{2}}^+$, and arguably the
${\textstyle\frac{5}{2}}^-$, can be discerned from the background. Low
energy compound system resonances from both calculations are at
essentially the same energies. Though as energy increases, the
centroid energies found from the (target state) width calculation
become increasingly higher than that of the no-width calculation.

Table~\ref{Be9-spec} summarizes results. For each spin-parity (column
1), the listed experimental $^9$Be resonance widths are given in
column 2. They are compared with predicted values for neutron emission
from the no-width (column 3) and width (column 5) calculations. In
columns 4 and 6 the ratios of the calculated widths to experiment are
given.  Boldface entries highlight the matches we find between theory
and experiment to within a factor of 3.
\begin{table}[h]\centering
\caption{\label{Be9-spec} Widths of states in $^9$Be 
 state widths ($\Gamma$ in MeV). In order, the columns designate the state
 spin-parity, the experimental values, the MCAS no-width calculation
 results,
 ratios ${\Gamma_\text{th.}}:{\Gamma_\text{exp.}}$, the MCAS width 
 calculation results, and ratios
  ${\Gamma_\text{th.}}:{\Gamma_\text{exp.}}$.}
\begin{tabular}{cc|cc|cc}
\hline
\hline
$J^\pi$ & $\Gamma_{exp.}$ & $\Gamma_{no-width}$ 
& ${\textstyle\frac{\Gamma_{no-width}}{\Gamma_{exp.}}}$ 
& $\Gamma_{width}$
& ${\textstyle\frac{\Gamma_{width}}{\Gamma_{exp.}}}$\\ 
\hline
${\textstyle\frac{1}{2}}^+$ & 0.217$\pm$0.001     & --- & ---  & 1.595 & 7.350\\
${\textstyle\frac{7}{2}}^-$ & 1.210$\pm$0.230     & 2.08$\times10^{-5}$ 
& 1.72$\times10^{-5}$ & 1.641 & \textbf{1.356}\\
${\textstyle\frac{1}{2}}^-$ & 1.080$\pm$0.110     & 0.495 & \textbf{0.458}  
& 1.686 & \textbf{1.561}\\
${\textstyle\frac{5}{2}}^+$ & 0.282$\pm$0.011     & 0.187 & \textbf{0.663}
& 0.740 & \textbf{2.624}\\
${\textstyle\frac{3}{2}}^-$ & 1.330$\pm$0.360     & 0.466 & \textbf{0.350}
& 3.109 & \textbf{2.337}\\
${\textstyle\frac{5}{2}}^-$ & $7.8{\times}10^{-4}$ & 0.060 & 76.74
& 2.772 & 3554\\
${\textstyle\frac{9}{2}}^+$ & 1.330$\pm$0.090     & 0.386 & 0.290
& 2.498 & \textbf{1.878}\\
${\textstyle\frac{3}{2}}^+$ & 0.743$\pm$0.055     & 3.286 & 4.423
& 5.162 & 6.947\\
\hline
\hline
\end{tabular}
\end{table}
Allowing the two excited states of $^8$Be to be resonances gives the
same spectral list as when they are treated as discrete, but the
evaluated widths of resonances in the compound nucleus significantly
increase. This is reflected in the cross sections. In the no-width
case, narrow and broad resonances are evident but most melt into the
background when the target states are allowed their widths. Where the
no-widths case resonances are often overly narrow, the width case
results are overly broad. While most of the theoretical $^9$Be
resonance widths given from the width calculation are closer, often
significantly, to experimental values, a few are not.

\begin{figure}[h]
\begin{center}
\scalebox{0.37}{\includegraphics*{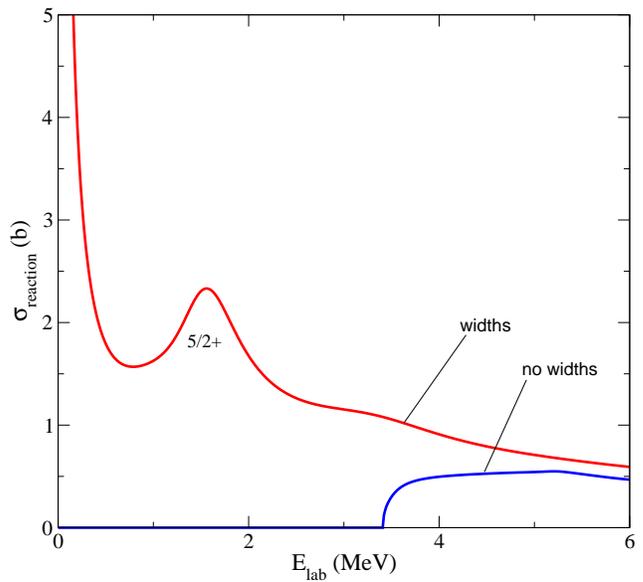}}
\end{center}
\caption{ \label{Be8+n-RC}
(Color online) Reaction cross sections for  
$n$-$^{8}$Be scattering from the two MCAS (widths and no widths)
calculations.}
\end{figure}

From Fig.~\ref{Be8+n} it seems that the cross section background decreases in
value with application of target state widths. The cause of this
reduction in cross section magnitude is the loss of flux inherent with
complex energies of the target states. That is illustrated by the 
reaction cross section of the $n + ^{8}$Be scattering,
shown in Fig.~\ref{Be8+n-RC}.
The reaction cross section found in the no-width case is zero until
3.4 MeV (3.03 in the centre of mass frame), the first inelastic
threshold.  However, for the width case, with the direct integration
method, there is a complex term in Eq.~(\ref{S-GgammaH}) for all
positive values of projectile energy.  Thus, the target state width
method cross section has flux loss from zero projectile energy
upwards. There is an asymptotic behaviour as energy approaches zero
(see Section~\ref{Sec-Theory}). While this could be attributed to the
presence of the ${\textstyle\frac{1}{2}}^+$ first resonance just above
threshold, this behaviour was also observed in the work of
Section~\ref{Sec-He6+n}, where no such low lying resonance exists.

In the reaction cross section from the width calculation, there is a
broad peak at ~1.5 MeV, corresponding to the
${\textstyle\frac{5}{2}}^+$ resonance.  This suggests that there is
more flux loss at energies corresponding to compound system
eigenstates than in the regions in between.  There is also a broad,
short bump at 3.4 MeV, but it is not clear how much loss of flux
arises from the opening of the second inelastic channel, as this
region also contains the theoretically broad ${\textstyle\frac{5}{2}}^-$
resonance.

\section{$p$+$^{8}$\protect\Be\: scattering}
\label{Sec-Be8+p}

As a further investigation into this phenomenon, we have used
nuclear interactions selected for neutrons on
$^8$Be (Table~\ref{Be8-param}) in calculations of the
scattering with protons. The Coulomb potential used is that of the
same deformed rotor and Woods-Saxon form, but with a Coulomb radius of 1.3 fm,
and a deformation, $\beta_2$, of 0.7.

\begin{figure}[h]
\begin{center}
\scalebox{0.53}{\includegraphics*{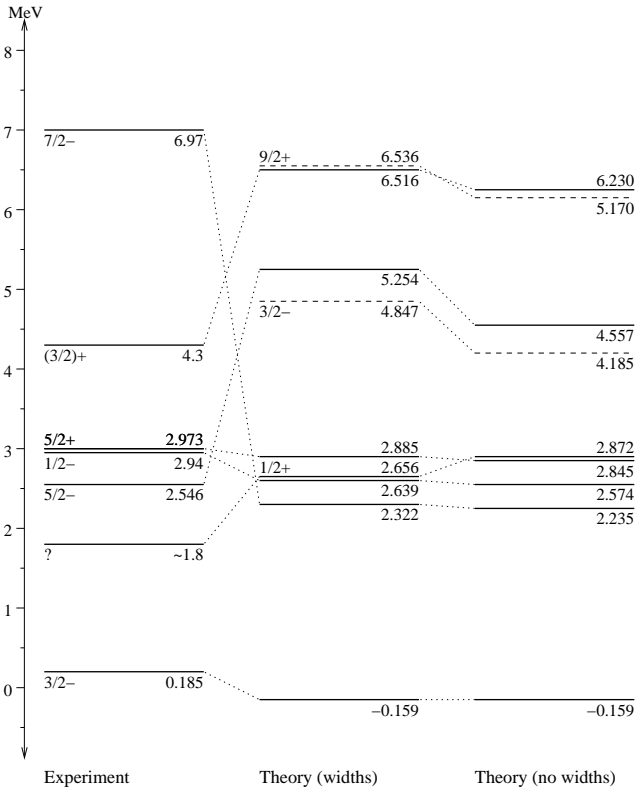}}
\end{center}
\caption{ \label{Be8+p-spec}
Experimental $^9$B spectrum and the (widths and no widths) 
MCAS results for the $p$+$^8$Be system.} 
\end{figure}

The experimental spectrum of the compound system, $^9$B, is compared
with theoretical spectra from  width and no-width calculations
as indicated in Fig.~\ref{Be8+p-spec}.
The observed states up to 8 MeV are matched by the calculated results, 
albeit in a few cases severely over- or under-bound. As the observed spectral
list replicates that of its mirror, $^9$Be, 
a spin-parity, $J^{\pi}$, of ${\textstyle\frac{1}{2}}^+$ 
is ascribed to the unknown state at $\sim$1.8 MeV excitation.
Likewise the unknown
state at 4.3 MeV can be assigned the $J^{\pi}$ of ${\textstyle\frac{3}{2}}^+$.
There are also two predicted states that are as yet unobserved.

As with the $n$+$^{8}$Be investigations, allowing the target states to
have widths alters the centroid energies of the compound system.
Again the centroids of the resonances are increased in comparison to
their counterpart found from the no-width calculation. This effect
increases with energy, with the exception of the state designated as
${\textstyle\frac{1}{2}}^+$.

Table~\ref{B9-spec} summarizes resonance width results for the
no-width and width calculations of $^9$B. The columns are as
designated in Table~\ref{Be9-spec}.  Dashed entries indicate
incomplete experimental data, and those marked N/A indicate states not
seen experimentally. Again the cases of agreement to within a factor
of 3 are highlighted.
\begin{table}[h]\centering
\caption{\label{B9-spec} Widths of $^9$B states ($\Gamma$ in MeV).
  In order the columns contain the spin-parity of the state, the
  the experimental width, those from a no-width MCAS  calculation,
  the ratios ${\Gamma_\text{th.}}:{\Gamma_\text{exp.}}$, 
  those from a width MCAS calculation, and ratios
  ${\Gamma_\text{th.}}:{\Gamma_\text{exp.}}$.}
\begin{tabular}{cc|cc|cc}
\hline
\hline
$J^\pi$ & $\Gamma_{exp.}$ & $\Gamma_{no-width}$ 
& ${\textstyle\frac{\Gamma_{no-width}}{\Gamma_{exp.}}}$ 
& $\Gamma_{width}$
& ${\textstyle\frac{\Gamma_{width}}{\Gamma_{exp.}}}$\\ 
\hline
${\textstyle\frac{7}{2}}^-$ & 2.0$\pm$0.20      & 1.56$\times10^{-4}$ 
& 7.78$\times10^{-4}$ & 1.645 & \textbf{0.82}\\
${\textstyle\frac{1}{2}}^-$ & 3.13$\pm$0.02     & 0.652 & 0.21  
& 2.087 & \textbf{0.67}\\
${\textstyle\frac{5}{2}}^+$ & 0.55$\pm$0.04     & 0.450 & \textbf{0.82} 
& 1.085 & \textbf{1.97}\\
${\textstyle\frac{1}{2}}^+$ & ---     & 6.186 & ---  
& 7.627 & ---\\
${\textstyle\frac{3}{2}}^-$ & N/A     & 1.133 & N/A 
& 5.074 & N/A \\
${\textstyle\frac{5}{2}}^-$ & 0.081$\pm$0.005 & 0.887 & 10.954 
& 5.024 & 62\\
${\textstyle\frac{9}{2}}^+$ & N/A     & 0.837 & N/A
& 3.169 & N/A\\
${\textstyle\frac{3}{2}}^+$ & 1.6$\pm$0.20     & 2.666 & \textbf{1.666}  
& 5.053 & 3.2\\
\hline
\hline
\end{tabular}
\end{table}
Application of target states with widths yields a better match between
the theoretical and experimental ${}^9$B resonance widths for the
${\textstyle\frac{7}{2}}^-$ and ${\textstyle\frac{1}{2}}^-$ cases, but
slightly worse results for the ${\textstyle\frac{5}{2}}^+$ and
${\textstyle\frac{3}{2}}^+$ resonances.  However, the
${\textstyle\frac{3}{2}}^+$ does not agree well with the known
centroid energy, as is the case with the most poorly
predicted resonance, the ${\textstyle\frac{5}{2}}^-$. Again
these are limitations such as one must expect on using a
simple model for the nuclear interactions.

\section{$p$+$^{14}$O scattering}  
\label{Sec-O14+p}

The low excitation spectrum of $^{14}$O has a $0^+;1$ ground state with a
small width for $\beta^+$ decay ($\tau = 70.606 \pm 0.018 s$),
followed by a band starting 0.545 MeV above the proton-${}^{13}$N
threshold and consisting of a $1^-;1$ state at
5.173 MeV, and then known proton-unstable resonances with spin-parities
of $0^+;1$ at 5.920 MeV (decay width of
$\le$0.05 MeV), of $3^-;1$ at 6.272 MeV (decay width of
0.103 MeV), and of $2^+;1$ at 6.590 MeV (decay width
of $\le$0.06 MeV)~\cite{Ti04}.

To investigate the sensitivity of scattering observables to such small
particle-decay widths, two evaluations of the $p + ^{14}$O resonance
spectrum were made.  Again, in both, the ground state was taken to be
zero-width.  Of the low energy spectrum, the $0^+_2$ and $2^+$ were
selected for coupling in a previous study, as in the mirror
$n$+$^{14}$C system they generated the negative low-energy parity
states in $^{15}$C~\cite{Ca06}. In the first calculation, both are
taken as zero-width (ignoring their known proton-decay widths) while
in the second, the upper limits of the known widths were used.  In
both calculations, the same nuclear interaction was used and it was
generated assuming the collective model with rotor character for the
structure.  The parameter values were chosen to be those used in a
previous paper~\cite{Ca06}, and as such are fitted for the no-width
case. For completeness, these are given in Table~\ref{O14-param}. In
this case, the proton distribution is taken as that of a uniformly
charged sphere of radius 3.1 fm.
\begin{table}\centering
\caption{\label{O14-param} 
The parameter values defining the nucleon $+$ mass-14 potential.}
\begin{supertabular}{>{\centering}p{28mm} p{27mm}<{\centering} 
p{27mm}<{\centering}}
\hline
\hline
 & Odd parity & Even parity\\
\hline
$V_{\rm central}$ (MeV) & -44.2 & -44.2\\
$V_{l l}$ (MeV)       & 0.42 & 0.42 \\
$V_{l s}$ (MeV)       & 7.0 & 7.0 \\
$V_{ss}$ (MeV)        & 0.0 & 0.0 \\
\end{supertabular}

\begin{supertabular}{>{\centering}p{19.5mm} p{21mm}<{\centering} 
>{\centering}p{21mm} p{19mm}<{\centering} }
\hline
\hline
Geometry & $R_0 = 3.1$ fm & $a = 0.65$ fm & $\beta_2 = -0.5$ \\
Coulomb &  $R_c = 3.1$ fm & $a_c =$ --- & $\beta_2 = 0.0$ \\ 
\end{supertabular}
\begin{supertabular}{>{\centering}p{21mm} p{20mm}<{\centering} 
p{20mm}<{\centering} p{20mm}<{\centering}}
\hline
\hline
                        &$0s_{1/2}$ &$0p_{3/2}$ &$0p_{1/2}$\\
\hline
 $0^+ \; \lambda^{(OPP)}$ & 1000 & 1000 & 1000\\
 $2^+ \; \lambda^{(OPP)}$ & 1000 & 1000 & 3.11\\
 $4^+ \; \lambda^{(OPP)}$ & 1000 & 1000 & 3.87\\
\hline
\hline
\end{supertabular}
\end{table}

\begin{figure}[h]
\begin{center}
\scalebox{0.53}{\includegraphics*{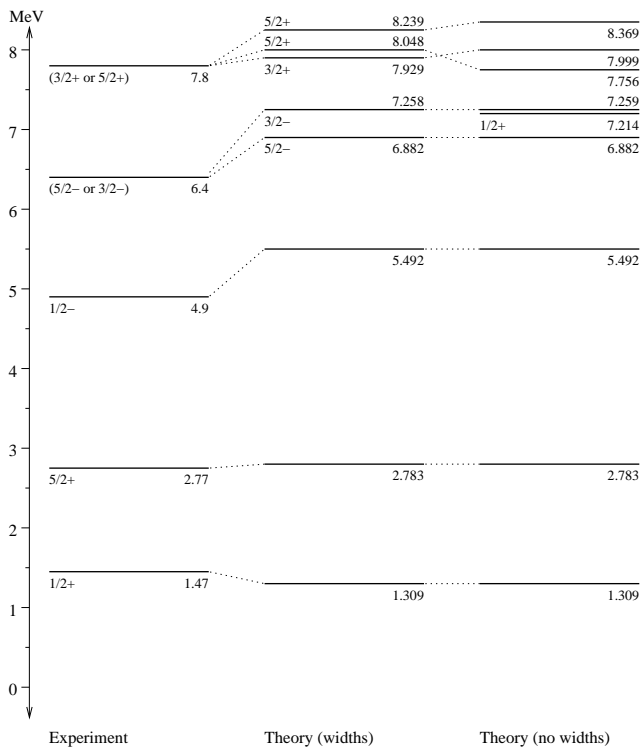}}
\end{center}
\caption{ \label{O14+p-spec}
Experimental $^{15}$F spectrum and those from MCAS
(widths and no widths) calculations  of the $p$+$^{14}$O system.}
\end{figure}

The compound system, ${}^{15}$F is particle unstable and only its
ground and first excited resonance states were known until
recently~\cite{Mu09}. Definite spin-parity assignments have been made
for the lowest three experimentally known states while the fourth and
fifth given in Ref.~\cite{Mu09}, tentatively are ${\textstyle\frac{3}{2}}^-$ or
${\textstyle\frac{5}{2}}^-$ for the 6.4 MeV state and 
${\textstyle\frac{3}{2}}^+$ or
${\textstyle\frac{5}{2}}^+$ for the 7.8 MeV state.  These resonance state
centroids are compared with the spectra predicted from our MCAS
calculations in Fig.~\ref{O14+p-spec}. The measured states pair up
quite reasonably with the calculated ones.  There is a close agreement
with the centroids of the theoretical resonances save for one.  Why
the ${\textstyle\frac{1}{2}}^+_2$ of the no-width calculation has no matching
partner in the spectrum from the width calculation is unknown at this
time. It should be noted that the parameters that produced this
spectrum were fitted only to the observed ${\textstyle\frac{1}{2}}^+$ and
${\textstyle\frac{5}{2}}^+$ states, before the results of Ref.~\cite{Mu09} were
published.

The ${}^{15}$F resonance width results for the no-width and width
calculations are in comparison with the known values in
Table~\ref{F15-spec}. The columns are as per
Tables~\ref{Be9-spec}.  Both calculations give good agreement (to
within the factor of 3 used as a measure) as is indicated by
favourable ratios being displayed in boldface.
\begin{table}[h]\centering
\caption{\label{F15-spec} $^{15}$F state widths ($\Gamma$
  in MeV) from experiment, calculation with $^{14}$O states taken as 
  zero-width, ratios ${\Gamma_\text{th.}}:{\Gamma_\text{exp.}}$, calculation
  with $^{14}$O states taken with known width, and ratios
  ${\Gamma_\text{th.}}:{\Gamma_\text{exp.}}$.}
\begin{tabular}{cc|cc|cc}
\hline
\hline
$J^\pi$ & $\Gamma_{exp.}$ & $\Gamma_{no-width}$ 
& ${\textstyle\frac{\Gamma_{no-width}}{\Gamma_{exp.}}}$ 
& $\Gamma_{width}$
& ${\textstyle\frac{\Gamma_{width}}{\Gamma_{exp.}}}$\\ 
\hline
${\textstyle\frac{1}{2}}^+$ & 1.00$\pm$0.20 & 0.77 & \textbf{0.77} & 0.77 & 
\textbf{0.77}\\ 
${\textstyle\frac{5}{2}}^+$ & 0.24$\pm$0.03 & 0.30 & \textbf{1.25} & 0.32 &
\textbf{1.33}\\
${\textstyle\frac{1}{2}}^-$   & 0.2$\pm$0.20 & 4.22$\times10^{-3}$ & 47 
& 5.43$\times10^{-2}$ & 3.68\\ 
(${\textstyle\frac{5}{2}}^-$) & 0.2$\pm$0.20 & 0.01  & 0.05 & 0.07 & \textbf{0.35}\\ 
(${\textstyle\frac{3}{2}}^-$) & 0.2$\pm$0.20 & 3.68$\times10^{-2}$  & 0.18 & 0.09 
& \textbf{0.45}\\
(${\textstyle\frac{3}{2}}^+$) & 0.4$\pm$0.40 & 3.67  & 9.19 & 0.16 & \textbf{0.4}\\ 
(${\textstyle\frac{5}{2}}^+$) & 0.4$\pm$0.40 & 0.42  & \textbf{1.05} & 0.07 & 0.18\\
(${\textstyle\frac{5}{2}}^+$) & 0.4$\pm$0.40 & 0.55  & \textbf{1.38} & 0.24 
& \textbf{0.61}\\
\hline
\hline
\end{tabular}
\end{table}

The two predicted resonances that may correspond to the fourth
observed state increase in width when target state widths are
introduced, bringing them closer to the experimental value in
Ref.~\cite{Mu09}, in line with comments in that paper.
However, the three predicted resonances that may correspond to the
fifth observed state decrease in width when target state widths are
introduced, bringing the ${\textstyle\frac{3}{2}}^+$ resonance closer to the
observed value, but taking the second of the two ${\textstyle\frac{5}{2}}^+$
resonances outside of the factor of 3 used as a measure of quality.

\begin{figure}[h]
\begin{center}
\scalebox{0.37}{\includegraphics*{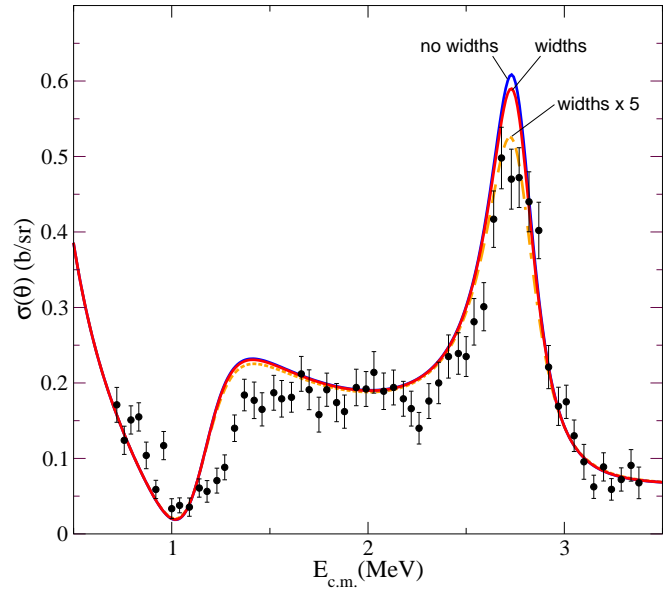}}
\end{center}
\caption{ \label{O14+p-147} (Color online) MCAS calculated elastic cross
  sections for $p$+$^{14}$O scattering at $147^\circ$ scattering angle 
  compared with experimental data~\cite{Go04}.
 Details are given in the text.}
\end{figure}

More telling, however, is the comparisons of calculated results with
$p + ^{14}$O cross sections.  Scattering of low energy ${}^{14}$O 
ions from hydrogen targets has been measured~\cite{Go04}, and a
comparison of our calculated cross sections with their data taken at a 
scattering angle of 147$^{\circ}$ is made in Fig.~\ref{O14+p-147}.

\begin{figure}[h]
\begin{center}
\scalebox{0.37}{\includegraphics*{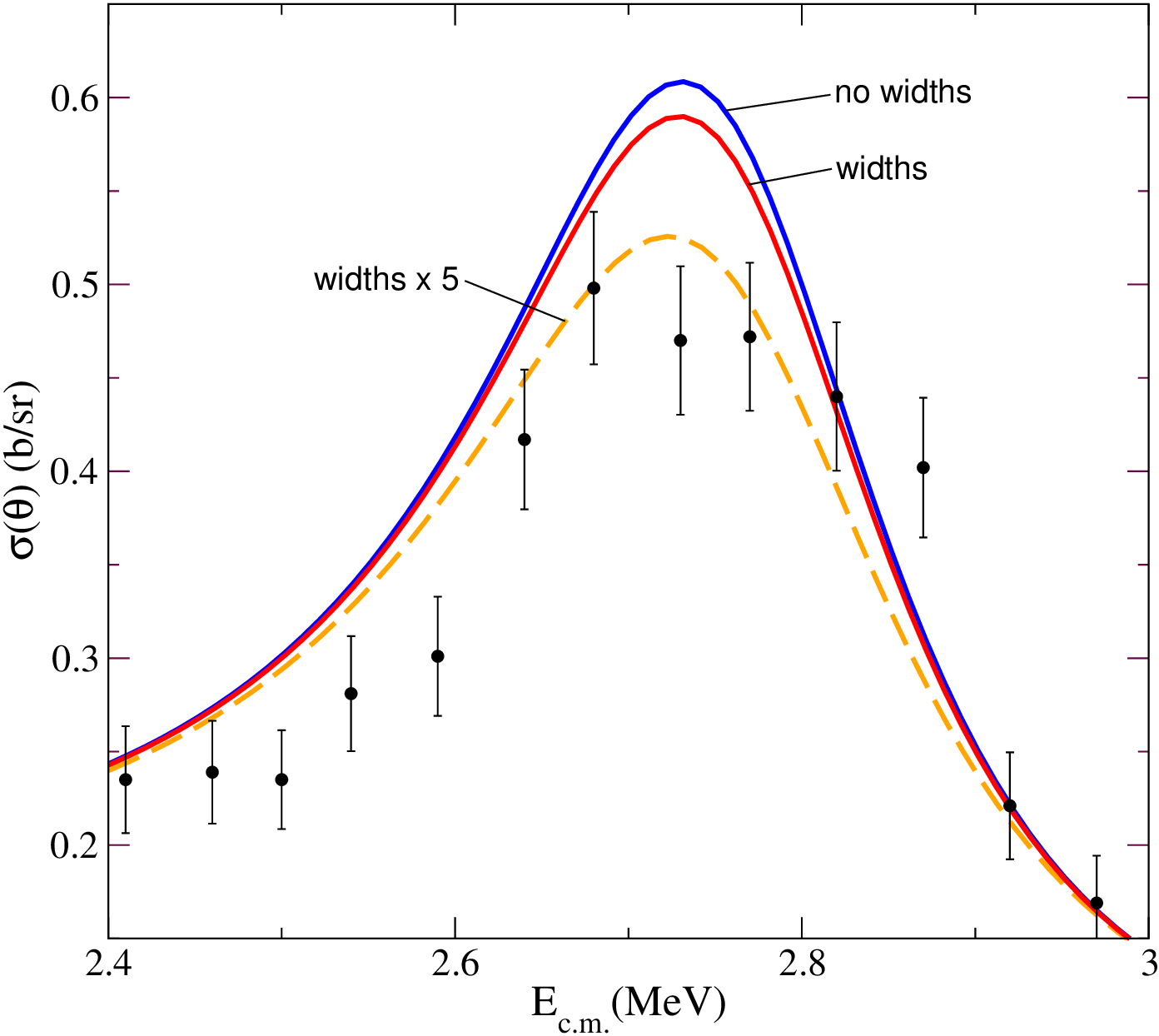}}
\end{center}
\caption{ \label{O14+p-147a}(Color online) Calculated elastic cross
  sections for $p$+$^{14}$O scattering 
  at 147$^\circ$ scattering angle for energies
  centred on the ${\textstyle\frac{5}{2}}^+$ resonance compared with the
 experimental data~\cite{Go04}.}
\end{figure}

\begin{figure}[h]
\begin{center}
\scalebox{0.37}{\includegraphics*{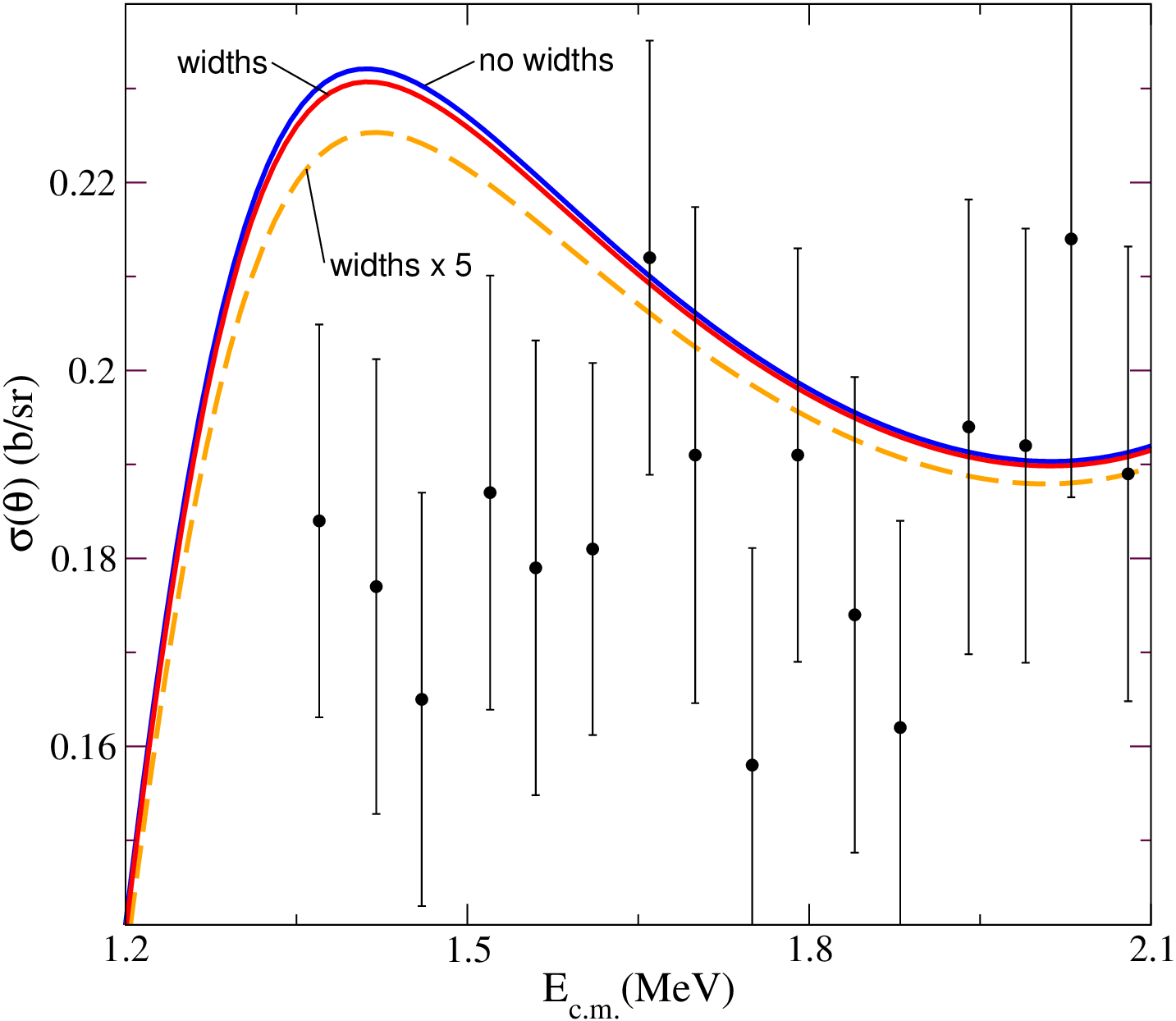}}
\end{center}
\caption{ \label{O14+p-147b}(Color online) Calculated elastic cross
  sections for $p$+$^{14}$O scattering 
  at 147$^\circ$ scattering angle for energies
  centred on the ${\textstyle\frac{1}{2}}^+$ resonance compared with the
 experimental data~\cite{Go04}.}
\end{figure}

Clearly both calculations give very good results, with obvious
evidence of the effect of the two resonances in the compound system.
Inclusion of experimental decay widths of the target states, in this
case, does not obviously improve the comparison with the data in this
figure. However, there is a little improvement in the energy variation
of the results over the ${}^{15}$F resonance regions, where the
no-width calculations had larger value than experiment.  The cross
sections for just those energy regions are shown in
Figs.~\ref{O14+p-147a} and \ref{O14+p-147b}. The improvements are
slight but in the correct direction. For comparison, a cross section
for a hypothetical case in which the target state widths are five
times larger than observed is displayed by the dashed line in
Figs.~\ref{O14+p-147}, \ref{O14+p-147a}, and \ref{O14+p-147b}.  At
this stage, experimental data of higher accuracy is most desirable, to
ascertain these fine details.

\section{$p$+$^{6}$\protect\He\:  scattering}
\label{Sec-He6+p}

To study the effect of target state particle-emission widths on
calculated resonance centroids over a large range of projectile
energies, the $p$+$^{6}$He system is examined. The known spectrum of
the compound system, $^7$Li, is particularly well suited to this as it
has known states over a 15~MeV interval.  As most of these compound
system states lie below the $p$+$^6$He threshold, the Heaviside
function discussed in Section~\ref{Sec-Theory} has not been used. The
aim here is not to produce cross sections, but to study of effects of
target state widths on the compound nucleus spectrum.

The low excitation spectrum of $^{6}$He consists of a $0^+;1$ ground
state with a very small width ($\tau = 806.7 \pm 1.5 ms$) for
$\beta^-$ emission, followed by a $2^+;1$ state at 1.797 MeV of width
0.113$\pm$0.02 MeV which emits neutrons and $\alpha$-particles,
followed, at 5.6 MeV, by a degenerate triplet designated
$(2^+,1^-,0^+);1$ with a width of 12.1$\pm$1.1 MeV. The decay paths of this
triplet are unknown, though particle instability can be assumed.

\begin{figure}[h]
\begin{center}
\scalebox{0.51}{\includegraphics*{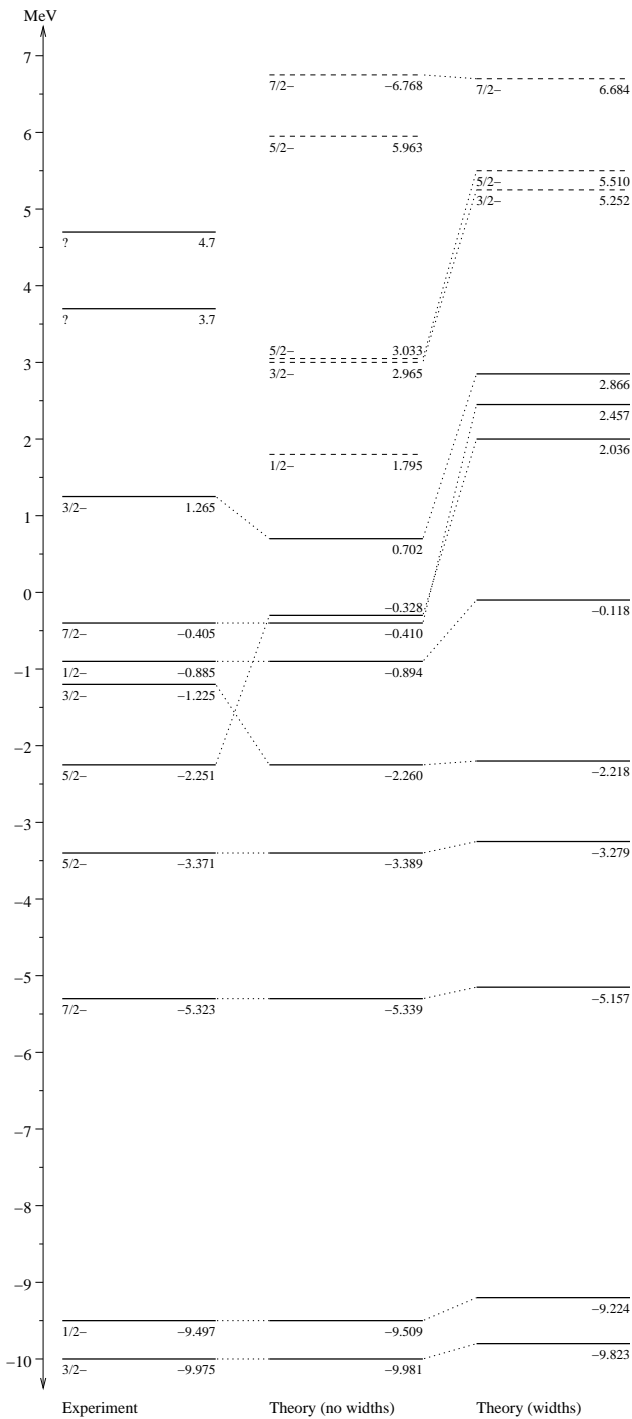}}
\end{center}
\caption{ \label{He6+p-spec}
Experimental $^7$Li spectrum and those from (widths and no widths) 
 MCAS calculations of the $p$+$^6$He system.}
\end{figure}

Two evaluations of the $p$+$^6$He cross section were obtained with
MCAS; in both, the ground state was taken as having zero-width.  In
the first calculation the excited states were also assumed to have
zero-width. In the second, the true width of the $2^+_1$ evaluated
from reaction data was used, and the full width of the triplet,
$(2^+,1^-,0^+);1$ was ascribed to the $2^+_2$ state. In both
calculations, the same nuclear interaction was considered, again taken
from a collective model with rotor character~\cite{Am03} using the
parameter values of Ref.~\cite{Ca06a}. They were determined by a
no-width calculation of the ${}^7$Li spectrum.  For completeness, they
are listed in Table~\ref{He6-param}.  The known compound-system
spectrum~\cite{Ti04} contains only negative parity states, and thus
there is no guide for fitting positive-parity parameters.  Thus,
positive parity interactions are not considered.  The proton
distribution was taken as a uniform charged sphere.
\begin{table}\centering
\caption
{\label{He6-param} 
The parameter values defining the trial base potential.}
\begin{supertabular}{>{\centering}p{28mm} p{27mm}<{\centering} 
p{27mm}<{\centering}}
\hline
\hline
 & Odd parity & Even parity\\
\hline
$V_{\rm central}$ (MeV) & -36.8177 & --- \\
$V_{l l}$ (MeV)        & -1.2346  & --- \\
$V_{l s}$ (MeV)        & 14.9618  & --- \\
$V_{ss}$ (MeV)         & 0.8511   & --- \\
\end{supertabular}
\begin{supertabular}{>{\centering}p{19.5mm} p{21mm}<{\centering} 
>{\centering}p{21mm} p{19mm}<{\centering} }
\hline
\hline
Geometry & $R_0 = 2.8$ fm & $a = 0.88917$ fm & $\beta_2 = 0.7298$ \\ 
Coulomb &  $R_c = 2.0$ fm & $a_c =$ --- & $\beta_2 = 0.0$ \\
\end{supertabular}
\begin{supertabular}{>{\centering}p{23mm} p{19mm}<{\centering} 
p{19mm}<{\centering} p{20mm}<{\centering}}
\hline
\hline
Proton blocking            &$0s_{1/2}$ &$0p_{3/2}$ &$0p_{1/2}$\\
\hline
 $0^+ \; \lambda^{(OPP)}$   & 1000 & 0.0 & 0.0\\
 $2^+_1 \; \lambda^{(OPP)}$ & 1000 & 0.0 & 0.0\\
 $2^+_2 \; \lambda^{(OPP)}$ & 1000 & 0.0 & 0.0\\

\hline
Neutron blocking           &$0s_{1/2}$ &$0p_{3/2}$ &$0p_{1/2}$\\
\hline
 $0^+ \; \lambda^{(OPP)}$   & 1000 & 17.8 & 36.0\\
 $2^+_1 \; \lambda^{(OPP)}$ & 1000 & 17.8 & 5.8\\
 $2^+_2 \; \lambda^{(OPP)}$ & 1000 & 17.8 & 5.8\\
\hline
\hline
\end{supertabular}
\end{table}
The actual known spectrum of the compound system, $^7$Li, and the
theoretical spectra found from our two MCAS evaluations
are displayed in Fig.~\ref{He6+p-spec}.  The reference energy is the 
$p$+$^6$He threshold and so all states below that energy 
in ${}^7$Li are discrete in so far as proton emission is concerned.
There are other thresholds for particle emission lying below this value
and so many are resonances in their own right.  For example
the $^3$He+$\alpha$ threshold is at 2.467 MeV so that the set of states
lying between the ${\textstyle\frac{7}{2}}^-$ ones are resonances with regard
to that break-up. MCAS studies have been made to note such~\cite{Ca06a}, 
though since the two nuclei involved have no excited states of import
that was a single channel study.

As the parameters are fitted to the no-width case, in this case, that
spectrum is shown next to the data. Note, these results for the
calculation without target state widths are marginally different from
those in Ref.~\cite{Ca06a}, by at most 18 eV, as the program suite has
undergone many small refinements since that publication.

The differences between these calculated spectra suggest that, as
projectile energy increases, the consideration of the target state
widths increasingly shifts the energy levels from the values found
from the no-width calculation.  Overall, the energy levels (and
resonance centroids) increase as the excitation energy increases
compared to the no-width results, similar to what was observed with
the $^8$Be investigations.  In addition, there are two resonances in
the spectrum found in the no-width case that are missing in the
results found from the width calculation. They are the
${\textstyle\frac{1}{2}}^-$ (1.795 MeV) and ${\textstyle\frac{5}{2}}^-$
(5.963 MeV)
resonances.  Also there is a ${\textstyle\frac{7}{2}}^-$ resonance (2.457 MeV)
only found in the width results.  Of the resonances above the
$p$+$^6$He threshold that MCAS predicts, only one is matched to an
observed state with definite $J^{\pi}$, the ${\textstyle\frac{3}{2}}^-$ at 1.265
MeV.  No comparison is drawn between the matches of predicted widths
to observation, as it is poorly recreated in centroid in both
calculation.

Note that, as with the ${\textstyle\frac{1}{2}}^+$ state in
Section~\ref{Sec-Be8+n}, the ${\textstyle\frac{5}{2}}_2^-$ and 
${\textstyle\frac{7}{2}}_2^-$
states are naturally moved from the negative to the positive energy
regime with the application of widths. Thus, the application of the
Heaviside function would mean there would be two representations of
this state; one in which it is a subthreshold state and the other in
which it is a resonance. However, as noted in Section~\ref{Sec-Be8+p},
occasionally the reverse occurs. This is an additional indication that
another method is required to prevent width-penetration into the
negative energy regime, one which localizes the effects of widths of
target states to the region surrounding their centroids.

Comparisons between the matches of predicted widths to observation for
the higher excitations are not good with either of the
calculations. Improved MCAS studies are needed.

\section{$n$+$^{6}$\protect\He\:  scattering}
\label{Sec-He6+n}

To investigate further the asymptotic increase in reaction cross
section as projectile energy approaches zero, the $n$+$^{6}$He system
was investigated. The nuclear interaction was set as that given in
Table~\ref{He6-param}, with neutron Pauli blocking weights and no
Coulomb interaction.

\begin{figure}[h]
\begin{center}
\scalebox{0.53}{\includegraphics*{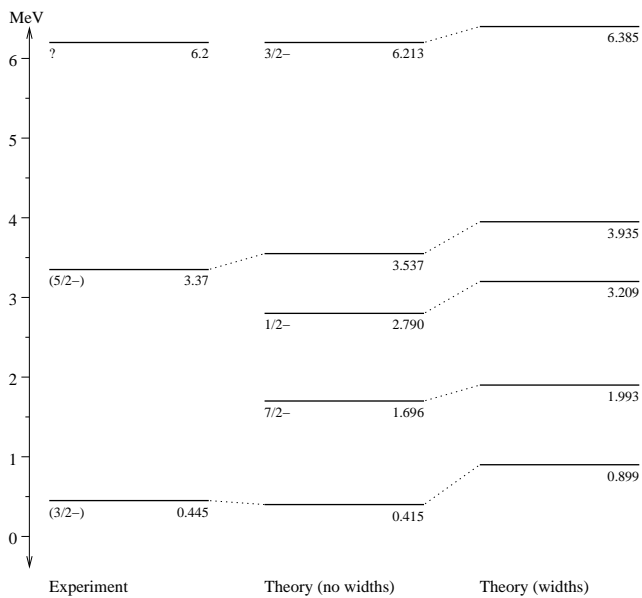}}
\end{center}
\caption{ \label{He6+n-spec}
Experimental $^7$He spectrum and those from 
(widths and no widths) MCAS calculations of the 
$n$+$^6$He system.}
\end{figure}

The experimental spectrum of the compound system, $^7$He, and
theoretical spectra where target state widths are considered and set
to zero are displayed in Fig.~\ref{He6+n-spec}. Again, as the
parameters are tuned to the no-widths case, this appears first after
experiment. Also, as no positive parity states are known, no positive
parity interaction was used in these calculations.

In this case there is much movement of resonance centroids when the
target states are taken to be resonances themselves. The trend,
though, is as found in the previous studies. Likewise the low energy
cross section (not depicted) has a similar asymptotic increase as
energy tends to zero as evident in the $n$+${}^8$Be case. This is
further evidence a more sophisticated method is required to prevent
width-penetration into the negative energy regime, one in which the
influence of target state widths approaches zero as projectile energy
approaches zero.

\section{Discussion and conclusions}
\label{conclus}

An extension to MCAS that considers particle-decay widths of target
nucleus eigenstates has been further tested for a range of light mass
nuclear targets.  A Heaviside step function has been applied to those
target state widths when used in the energy denominators of the
coupled-channel Lippmann-Schwinger equations. This ensures that the
Lorentzian shape of the target state resonances has no effect below
threshold ($E=0$) which would compromise stability in the compound
bound system.

Experimental cross section data at 147$^{\circ}$ for $p$+$^{14}$O
scattering has been compared with previously published results.
Inclusion of defined target state widths gives only a marginally
better representation of resonance magnitudes in the cross section.
Using target state widths five times the known size brings the larger
${\textstyle\frac{5}{2}}^+$ resonance into agreement with experiment.

In addition to evidence supporting previous findings that
consideration of target state widths increase the calculated widths of
elastic cross section resonances, and that at low energy, the centroids
of such are little altered, new trends have been found.  First, when
comparing results of calculations that include target state widths to
those from calculations that do not, the compound nucleus spectrum,
generally, will be wider spread with most centroids increasing with
excitation energy.  This phenomenon is evident in the investigations
of the $n$+$^8$Be, of the $p$+$^8$Be, of the $p$+$^6$He, and
especially of the $n$+$^6$He systems.  The effect seems to increase as
target state widths increase.

To ensure that any resonance aspect of target states does not have
influence below the nucleon-nucleus threshold in our formalism, we have 
used a Heaviside function in the specification of the coupled-channel
Green's function. This, however, is too simplistic.  A first problem
with the approach occurs with specification of compound 
states that lie near to the threshold energy.  In the case of the
${}^9$Be spectrum, a very weakly bound state was found with the no-width
calculation which became a resonance with very low centroid energy 
upon allowing the target states to be resonances with Lorentzian
character. The Heaviside weighting, however, retains the state as a
weakly bound one.   This problem is caused in the current formalism as 
the effect of a target width is felt for all positive values of projectile 
energy, the Lorentzian never having a zero value. A better
energy dependent scaling factor is needed; one with value one at the
centroid energy and  approaching zero for lower and higher energies
faster than a Lorentzian. Another problem with the current model
is that reaction cross sections increase asymptotically as the projectile 
energy tends to zero.  New investigations will see if that undesirable
effect can be offset with a more appropriate
energy dependent scaling factor applied to the target state

A further problem of the current formalism, one not observed 
previously~\cite{Fr08a}, is that at the larger excitation energies, 
use of target state widths can change the calculated spectrum 
of the compound system significantly. 
Our $p$+$^{14}$O and $p$+$^6$He studies showed that.
Hence when allowing target states to be resonances themselves, the nuclear 
interactions deemed best from a no-width study may need adjustment
in concert with the specification of target state resonances.

\begin{flushleft}
\section*{Acknowledgments}
\end{flushleft}
P.F. gratefully acknowledges support from the Instituto de Ciencias
Nucleares, UNAM.  Travel support came also from Natural Sciences and
Engineering Research Council, Canada, INFN, sez. di Padova, Italy, the
National Research Foundation, South Africa, and from the
International Exchange program of the Australian Academy of Science.

\bibliography{Fraser-RMF}

\end{document}